\documentclass[twocolumn,prl]{revtex4-1}
\usepackage{graphicx}
\usepackage{amsmath}
\usepackage{amssymb}
\usepackage{bm}

\begin{document}

\noindent {\bf Comment on \textquotedblleft Quantum Time Crystals\textquotedblright: a new paradigm or just another proposal of \emph{perpetuum mobile} ?}

\vspace*{0.5\baselineskip}

In a recent Letter \cite{Wilczek2012}, Wilczek proposes the existence of a new state of matter, \textquotedblleft quantum time crystals\textquotedblright, defined as systems which, in their quantum mechanical ground state, display a time-dependent behavior (periodic oscillation) of some physical observable. The proposal is based upon a model consisting of (discernible) particles on an Aharonov-Bohm (AB) ring, attractively coupled to each other by a contact interaction.
Using mean field theory, the problem reduces to the non-linear Schr\"{o}dinger equation (NLSE),
$\mathrm{i}\partial_t \psi\! =\! \left[\frac{1}{2} (-\mathrm{i}\partial_{\phi} - \alpha)^2 -\lambda |\psi|^2 \right]\psi$,
with periodic boundary condition
$\psi(\phi \!+\! 2\pi)\! =\! \psi(\phi)$, and $\int_0^{2\pi}\! \mathrm{d}\phi\,|\psi|^2=1$.
Wilczek first solves it for zero AB flux
($\alpha =0$):
non surprisingly, the system undergoes a phase transition towards formation of a lump (bright soliton) for a coupling strength larger than a threshold
($\lambda \geq \frac{\pi}{2}$).
Turning then to the case of non-zero flux $\alpha$, Wilczek constructs a solution of the NLSE by setting the zero-flux soliton state into rotation at angular velocity
$\omega=\alpha$
(so that the apparent flux vanishes in the rotating frame); this solution has an energy
$\Delta\epsilon=\frac{\alpha^2}{2}$
per particle above the zero-flux ground state (I restrict here the discussion to
$|\alpha| \leq \frac{1}{2}$,
which is sufficient, since all physical properties are periodic in $\alpha$, with period 1 \cite{ByersYang1961}). The physical interpretation of this result is clear: starting from the zero-flux ground state and ramping up the flux, the lump experiences a torque (from Faraday's law) and is accelerated to the angular velocity
$\omega=\alpha$.
The gained energy
$\Delta\epsilon$
is just the corresponding rotational kinetic energy. The crucial question then is: is this rotating-soliton solution the ground state of the system ? Wilczek answers \textquotedblleft yes\textquotedblright\ without further justification, and concludes that his model thus constitutes a \textquotedblleft quantum time crystal\textquotedblright. However, Wilczek did \emph{not} prove the absence of any lower-energy solution to the NLSE.

On the other hand, one can readily observe that Wilczek's result leads to paradoxical (unphysical) consequences. (i) Let us consider the large coupling limit
($\lambda \to +\infty$).
In that limit, the soliton width shrinks to zero like
$\lambda^{-1}$,
and the wavefunction amplitude near the antipode of the soliton shrinks exponentially
($|\psi| \! \sim \! \sqrt{\lambda}\, \mathrm{e}^{-\lambda\pi /2}$).
The sensitivity of the system to the AB flux $\alpha$ should also be exponentially small (in particular, the flux-induced variation of the ground state energy should be exponentially small as well), and the dynamics of a classical lump (which is of course completely insensitive to the AB flux and has a static ground state) should be recovered in the limit
$\lambda \to +\infty$,
in striking contrast with Wilczek's result. (ii) When coupled to some external environment (e.g., the electro-magnetic field, if the particles are considered to carry some electric charge), the rotating lump would radiate energy while being in its ground state, thereby violating the principle of energy conservation (arguably physics' strongest principle). Wilczek's considerations on this highly critical issue, namely the suggestion that the coupling to the environment could be reduced by using higher multipoles or suppressed by placing the system in a cavity, amount to dismissing the problem without addressing the paradox convincingly.

These remarks thus strongly suggest that Wilczek's rotating-soliton state is \emph{not} the ground state and that the true ground state is actually a stationary state, as I show below. The solution of the NLSE for arbitrary flux is too lengthy and technical to fit in this Comment (the reader is referred to Ref.~\cite{Kanamoto2003} for details); thus I shall give here only the solution for
$\alpha =\frac{1}{2}$,
which is sufficient to disprove Wilczek's claim. One first notices that the flux $\alpha$ can be gauged away from the NLSE by the transformation
$\psi(\phi)\! =\! \mathrm{e}^{\mathrm{i}\alpha\phi}\tilde{\psi}(\phi)$,
resulting in the twisted boundary condition,
$\tilde{\psi}(\phi\! + \! 2\pi)=\mathrm{e}^{-\mathrm{i}2\pi \alpha}\tilde{\psi}(\phi)$.
So, for
$\alpha=\frac{1}{2}$,
one simply has to solve the NLSE with
$\alpha\!\equiv \! 0$
and antiperiodic boundary condition. The correct ground state has the following stationary wavefunction:
$\tilde{\psi}(\phi) = \frac{kK}{\pi\sqrt{\lambda}}\, \mathrm{cn}(\frac{\phi K}{\pi},k)$;
$K\equiv \mathrm{K}(k)$ and $E\equiv \mathrm{E}(k)$
are the complete elliptic integrals of first and second kind, and
$\mathrm{cn}(u,k)$
is a Jacobi elliptic function \cite{NIST2010}; the elliptic modulus $k$ satisfies
$[E\! -\! (1 \! - \! k^2)K]K \! =\! \frac{\pi\lambda}{2}$.
The chemical potential is
$\mu\! =\! \frac{K^2}{\pi^2}\left(\frac{1}{2} \! -\! k^2\right)$
and the total energy per particle is
$\epsilon\! \equiv\! \mu\! +\! \frac{\lambda}{2} \int_0^{2\pi}\!\! \mathrm{d}\phi\, |\psi|^4 \!=\! - \frac{K^2 [(2k^2-1)E - (1-k^2)(3k^2-1)K] }{6\pi^2 [E-(1-k^2)K]} $ \cite{note}.
Solving explicitly these equations confirms that the present state has a lower energy than Wilczek's one. For strong coupling
($\lambda\to +\infty$),
fully analytical results can be obtained for any value of the AB flux $\alpha$: the flux dependence of ground state energy then takes the simple asymptotic form
$\Delta\epsilon\! = \! -3 [1 \! - \! \mathrm{cos}(2\pi\alpha )] \lambda^2 \mathrm{e}^{-\pi\lambda}$,
which is in fact negative (this is due to the lump being narrower for
$\alpha=\frac{1}{2}$ than for $\alpha=0$,
leading to more effective attractive coupling) and much lower than Wilczek's result
($\Delta\epsilon\! =\! \frac{\alpha^2}{2}$),
and does not lead to any unphysical paradox.

Wilczek himself admitted that his proposal is \textquotedblleft perilously close to fitting the definition of a \emph{perpetuum mobile}\textquotedblright \cite{Wilczek2012}. In the light of the above discussion, it seems that the very existence of \textquotedblleft quantum time crystals\textquotedblright\ remains highly speculative.

I am grateful to Andres Cano and Efim Kats for helpful comments and discussions.

\vspace*{0.5\baselineskip}

\noindent Patrick Bruno\\
{\small European Synchrotron Radiation Facility, BP 220, F-38043 Grenoble Cedex, France} \\
{ } \\

\end{document}